\documentstyle[12pt,lscape]{article}
\begin{document}

\begin{center}
{\Large
Analysis of $p$C-interactions at Momentum of 4.2 GeV/$c$ Within
Framework of FRITIOF and Cascade Models }
\end{center}

\vspace{1cm}
\begin{center}
A.S.Galoyan, E.N.Kladnitskaya, O.V.Rogachevskii, R.Togoo, V.V.Uzhinskii
\end{center}
\vspace{1cm}

\centerline{Abstract}
\vspace{0.5cm}
\noindent Experimental data on multiplicities and kinematical
characteristics of $\pi^-$, $\pi^+$ mesons and protons in the
interactions of protons with carbon nucleus at momentum 4.2 GeV/$c$ in
a dependence of collision centrality are analyzed. Parameter $Q$
which is a difference between multiplicities of positive and negative
charged particles without multiplicity of slow protons with momentum
less than 0.3 GeV/$c$ in an event, is taken as a criteria of collision
centrality. The experimental data on events with different centrality
are compared with predictions of the cascade-evaporation model and
the modified FRITIOF model.

It is shown that the cascade model does not reproduce decrease
of the average transverse momenta of participating protons
with increase of the centrality. The model overestimates the yield of
the particles in the target fragmentation region.

For the first time, non-nucleonic degrees of freedom in
nuclei ($\Delta ^+, \Delta ^0 $ isobars) are taken into account in the
FRITIOF model, and a commonly good description of the secondary
particles characteristics is reached.

\newpage

A large volume of experimental data on hA- and
AA-interactions at the momentum of 4.2 GeV/c has been obtained with the
help of 2-meter propane bubble chamber of the laboratory of High Energy
of JINR. The different theoretical models: Cascade Evaporation Model
(CEM)\cite{CEM_1,CEM_6}, Quark-Gluon String Model (QGSM)\cite{QGSM},
and FRITIOF model \cite{FRITIOF} were used for
the experimental data analysis. It was shown that the CEM describes
quite well the main characteristics of proton-carbon ($p$C) interactions.
For carbon-carbon (CC) interactions, the CEM overestimates the produced
particle multiplicity. The situation becomes worse at a description of the
characteristics of interactions with heavy nuclei, for example, A+Ta
collisions. The study of the interactions with heavy nuclei is
interesting for various applied tasks such us solving of the problems
connected with the creation of the subcritical nuclear reactors driven
by accelerator.
Statistics of the interactions with heavy nuclei is small as a rule,
and methodical corrections due to absorption of produced particles in
target are large. But one can study the multi-nucleon
interactions dominated in the interactions with the heavy nuclei using
the data about interactions with light nuclei, in particular
$p$C-interaction at different centrality of proton
collision with carbon nuclei presented in Ref. \cite{pC}.

At the experimental study $\pi^+$, $\pi^-$-mesons, participating
protons at the momentum $p> 0.3$ GeV/c and evaporated protons at the momentum
$0.15 \leq p\leq 0.3$ GeV/c were considered. Two groups of the protons
were distinguished: protons at the momentum from 0.3 to 0.75 GeV/c (these
are basically proton-participants from target nuclei), and the protons
at the momentum more than 0.75 GeV/c. The last group consists of
projectile protons interacted with the target nucleus and part of the protons
of the carbon nuclei obtained large transverse momentum at the interaction.
The average characteristics of the produced particles are given in the
tables 1, 2.


The parameter Q was accepted as a measure of collision centrality of
the $p$C-interactions. It was defined as: $Q=n_+ - n_- -n_{p.ev}$, where
$n_+$ and $n_-$ were the multiplicities of positive and negative charged
particles, correspondingly, and $n_{p.ev}$ was the multiplicity of
the evaporated protons. The value of Q is equal to total charge of the
particles taking an active part in the interaction. It correlates with
impact parameter magnitude. The value of centrality Q increases with
decrease of the impact parameter.

The table 1 gives the number of the analyzed $p$C-events and the
average multiplicities of secondary particles for all $p$C-interactions
and for six group of events at Q=1, 2, 3, 4, 5, $\geq6$.
One can see, the peripheral interactions (Q$\leq 2$) presents more than 70
\% of all inelastic $p$C-collisions. The part of most central
interactions (Q more than 4) is small and is about few percents.
As consequence, the all $p$C-interactions are characterized by the
average number of the participating protons $<n_p^{part}>$ less than 2.
The average multiplicity of $\pi^+$- mesons exceeds considerably the average
multiplicity of $\pi^-$-mesons. That is typical for proton interactions
with symmetrical nuclei with $N_p=N_n$.

We use the CEM \cite{Zhenis} and
two versions of modified FRITIOF model \cite{Adamovich, Uzhi} for
analysis of the experimental data. In the modified model FRITIOF,
it is assumed \cite{Khaled}  inelastic interaction of
projectile nucleon and target nucleon initiates reggeon exchanges
between spectator nucleons of the nucleus. In the cascade model, these
exchanges are interpreted as NN-collisions. We have used two variants
of the FRITIOF model. In the first version (DFRITIOF), it was
considered  the part of the nucleons knocked out by the reggeon
cascade are emitted as $\Delta^0-$ and $\Delta^+$-isobars. In the other
variant of the model, the $\Delta$-isobars in the spectator part of
nucleus were not taken into account.

\newpage
\begin{landscape}
\vspace{-7cm}

\begin{table}
\caption{The average multiplicities of the particles in the
$p$C-interactions at the momentum of 4.2 GeV/c at the different collision
centralities, e - experiment \protect \cite{pC},
m - the FRITIOF model calculations with $\Delta $-isobars.}

\bigskip

\begin{tabular}{|cr|c|c|c|c|c|c|c|} \hline
&& & & & & & & \\
{Q}     &       &       1          &        2        &        3        &        4        &       5         &         6       & {\mbox all events}\\
&& & & & & & & \\
\hline
$N_{ev}$ (\%)            &e &  2289 (27.3)    &  3814 (45.6)    &  1477 (17.6)    &  575 (6.9)      &  164 (1.9)      &   52 (0.62)     &   8371 (100)        \\
                          &m & 28457 (28.4)    & 37635 (37.6)    & 16675 (16.7)    & 9551 (9.6)      & 5166 (5.2)      & 2516 (2.5)      & 100000 (100)        \\   \hline
$<n_\pm>$                 &e & 2.72 $\pm$0.08  & 3.15 $\pm$0.02  & 4.697$\pm$0.04  & 5.73 $\pm$0.07  & 6.72 $\pm$0.12  & 7.60 $\pm$0.20  & 3.61 $\pm$0.02      \\
                          &m & 2.152$\pm$0.008 & 2.926$\pm$0.007 & 4.594$\pm$0.014 & 6.00 $\pm$0.02  & 6.96 $\pm$0.02  & 7.71 $\pm$0.03  & 3.627$\pm$0.007     \\   \hline
$<n_{\pi^-}>$             &e & 0.522$\pm$0.013 & 0.321$\pm$0.007 & 0.423$\pm$0.016 & 0.476$\pm$0.027 & 0.43 $\pm$0.05  & 0.36 $\pm$0.07  & 0.407$\pm$0.006     \\
                          &m & 0.479$\pm$0.004 & 0.321$\pm$0.003 & 0.424$\pm$0.005 & 0.448$\pm$0.006 & 0.45 $\pm$0.01  & 0.46 $\pm$0.01  & 0.406$\pm$0.002     \\   \hline
$<n_{\pi^+}>$             &e & 0.416$\pm$0.010 & 0.660$\pm$0.008 & 0.965$\pm$0.020 & 1.22 $\pm$0.04  & 1.40 $\pm$0.08  & 1.58 $\pm$0.16  & 0.706$\pm$0.007     \\
                          &m & 0.379$\pm$0.003 & 0.662$\pm$0.004 & 0.787$\pm$0.006 & 0.857$\pm$0.008 & 0.89 $\pm$0.01  & 0.93 $\pm$0.02  & 0.640$\pm$0.002     \\   \hline
$<n_p^{part}>$            &e & 1.054$\pm$0.015 & 1.743$\pm$0.010 & 2.526$\pm$0.024 & 3.22 $\pm$0.04  & 4.02 $\pm$0.09  & 5.10 $\pm$0.18  & 1.860$\pm$0.010     \\
                          &m & 1.088$\pm$0.005 & 1.658$\pm$0.004 & 2.624$\pm$0.007 & 3.54 $\pm$0.01  & 4.46 $\pm$0.02  & 5.75 $\pm$0.03  & 2.085$\pm$0.004     \\   \hline
$<n_p^{part}>$            &e & 0.241$\pm$0.009 & 0.584$\pm$0.009 & 1.212$\pm$0.024 & 1.84 $\pm$0.05  & 2.61 $\pm$0.10  & 3.39 $\pm$0.21  & 0.747$\pm$0.009     \\
$0.3<P\leq 0.75$ (GeV/$c$)&m & 0.114$\pm$0.002 & 0.454$\pm$0.003 & 1.219$\pm$0.006 & 2.03 $\pm$0.01  & 2.89 $\pm$0.02  & 4.17 $\pm$0.03  & 0.855$\pm$0.004     \\   \hline
$<n_p^{part}>$             &e & 0.588$\pm$0.020 & 0.740$\pm$0.018 & 0.664$\pm$0.027 & 0.57 $\pm$0.04  & 0.47 $\pm$0.06  & 0.56 $\pm$0.11  & 0.668$\pm$0.01     \\
$P>1,4$ (GeV/$c$)         &m & 0.785$\pm$0.006 & 0.794$\pm$0.005 & 0.712$\pm$0.007 & 0.62 $\pm$0.01  & 0.54 $\pm$0.01  & 0.44 $\pm$0.01  & 0.739$\pm$0.003     \\   \hline

$<n_p^{ev}>$              &e & 0.732$\pm$0.020 & 0.425$\pm$0.013 & 0.779$\pm$0.026 & 0.82 $\pm$0.04  & 0.87 $\pm$0.03  & 0.56 $\pm$0.08  & 0.640$\pm$0.009     \\
$0.15<P\leq 0.3$ (GeV/$c$) &m & 0.206$\pm$0.004 & 0.284$\pm$0.004 & 0.759$\pm$0.009 & 1.15 $\pm$0.01  & 1.16 $\pm$0.01  & 0.57 $\pm$0.01  & 0.476$\pm$0.003     \\   \hline
$<n_p^{ev}>$              &e & 5.32$\pm$0.02 & 0.49$\pm$0.01 & 3.15$\pm$0.03 & 2.22 $\pm$0.05  & 1.15 $\pm$0.01  & 0.11 $\pm$0.15  & 4.20$\pm$0.02     \\
$P< 0.15$ (GeV/$c$)     &m & 5.80 0.$\pm$0.003 & 4.716$\pm$0.003 & 3.255$\pm$0.009 & 1.89 $\pm$0.01  & 0.94 $\pm$0.01  & 0.21 $\pm$0.01 & 4.204$\pm$0.006  \\   \hline
\end{tabular}
\end{table}
\end{landscape}

\begin{landscape}
\vspace{-5cm}
\begin{table}
\caption{ The average momenta and angles of $\pi $-mesons
in the $p$C-interactions at 4.2 GeV/c at the different $Q$,
e - experiment \protect \cite{pC}, m - the FRITIOF model calculations with
$\Delta $- izobars.}
 \label{tabl2}
 \hspace{-0.5cm}
{ 
\begin{center}
\begin{tabular}{|c|c|c|c|c|c|c|c|}
\hline
 & & & & & & & \\
Q  &  1  &  2  &  3  &  4  &  5  &  $\geq $6  & {\mbox all events} \\
 & & & & & & &
\\  \hline
\hspace{1.9cm} \vspace{-0.1cm} e &
   0.567$\pm$0.014 & 0.518$\pm$0.010  & 0.424$\pm$0.014 & 0.375 $\pm$0.018 &
   0.38$\pm$0.04   & 0.45$\pm$0.07    & 0.503$\pm$0.007 \\
 \vspace{-0.1cm}\hspace{-0.2cm}$<p_{\pi -}\hspace{-0.1cm}>$,GeV/c
 & & & & & & &\\
 \hspace{1.9cm}m
 & 0.496$\pm$0.003 & 0.449$\pm$0.003 & 0.378$\pm$0.003  & 0.333$\pm$ 0.003
 & 0.314$\pm$0.004 & 0.295$\pm$0.005 & 0.429$\pm$0.002 \\ \hline
 \hspace{1.9cm} \vspace{-0.1cm} e &
  246$\pm$0.005 & 0.255$\pm$0.004 &
 0.248$\pm$0.007 & 0.236$\pm$0.011 & 0.215$\pm$0.025 & 0.27$\pm$0.06 &
 0.248$\pm$0.003 \\
 \vspace{-0.1cm}\hspace{-0.2cm}$<p_t^{\pi -}\hspace{-0.1cm}>$,GeV/c &
 & & & & & &\\
 \hspace{1.9cm}m &
 0.241$\pm$0.001 & 0.222$\pm$0.001 & 0.214$\pm$0.001 & 0.207$\pm$ 0.002
 & 0.208$\pm$0.002 & 0.199$\pm$0.003 & 0.224$\pm$0.001 \\ \hline
 \hspace{1.9cm} \vspace{-0.1cm} e &
45.2$\pm$1.0 & 49.5$\pm$1.0 & 57.3$\pm$1.5 & 62.1$\pm$2.3 &
62.3$\pm$4.9  & 62.3$\pm$11.0 & 50.8$\pm$0.6 \\
 \vspace{-0.1cm}\hspace{-0.3cm}$<\theta _{\pi -}\hspace{-0.1cm}>$,grad
 & & & & & & &\\
 \hspace{1.9cm}m & 47.4$\pm$0.3 & 49.6$\pm$0.3 &
 56.4$\pm$ 0.4 & 60.9$\pm$0.6 & 63.1$\pm$0.8 &66.8$\pm$1.1 &
 52.4$\pm$0.2 \\ \hline \hspace{1.8cm} \vspace{-0.1cm} e &
 0.564$\pm$0.007 & 0.554$\pm$0.004 & 0.505$\pm$0.006 & 0.475$\pm$0.007
 & 0.430$\pm$0.012 & 0.446$\pm$0.020 & 0.528$\pm$0.003 \\
 \vspace{-0.1cm}\hspace{-0.2cm}$<p_{\pi +}\hspace{-0.1cm}>$,GeV/c &
 & & & & & &\\
\hspace{1.9cm}m &
   0.592$\pm$0.004 & 0.533$\pm$0.002 & 0.428$\pm$0.003 &
   0.373$\pm$0.003 & 0.337$\pm$0.002 & 0.311$\pm$0.004 & 0.480$\pm$0.001
 \\ \hline
 \hspace{1.9cm} \vspace{-0.1cm} e&
  0.239$\pm$0.002 & 0.269$\pm$0.002 & 0.275$\pm$0.003 & 0.265$\pm$0.004
& 0.267$\pm$0.007 & 0.30$\pm$0.012 & 0.265$\pm$0.001 \\
 \vspace{-0.1cm}\hspace{-0.2cm}$<p_t^{\pi +}\hspace{-0.1cm}>$,GeV/c &
 & & & & & &\\
 \hspace{1.9cm}m &
 0.238$\pm$0.001 & 0.242$\pm$0.001 & 0.229$\pm$0.001 & 0.217$\pm$0.001
& 0.209$\pm$0.002 & 0.203$\pm$0.002 & 0.232$\pm$0.001
 \\ \hline
 \hspace{1.9cm} \vspace{-0.1cm} e &
 39.1$\pm$0.4 & 47.7$\pm$0.3 & 55.3$\pm$0.5 & 57.4$\pm$0.7 &
 64.9$\pm$1.2 & 68.7$\pm$2.0 & 50.3$\pm$0.2 \\
 \vspace{-0.1cm}\hspace{-0.3cm}$<\theta _{\pi +}\hspace{-0.1cm}>$,grad &
 & & & & & &\\
 \hspace{1.9cm}m &
 38.2$\pm$0.3 & 44.0$\pm$0.2 & 51.4$\pm$0.3 & 55.2$\pm$0.4 &
 58.6$\pm$0.5 & 61.5$\pm$0.7 & 47.6$\pm$0.1
 \\ \hline
\end{tabular}
\end{center}
}
\end{table}
\end{landscape}

Fig. 1 gives presentation of multiplicity distributions of
the different types of the produced particles. The largest number of
charged particles, registered in $p$C-interactions, achieves 13, of
$\pi^+$- and $\pi^-$-mesons -- 4, and of the proton-participants -- 8
(with taking into account exchanges $p\rightarrow n$ and $n\rightarrow
p$). The points are the experimental data \cite{pC}, the solid lines are the
calculations by DFRITIOF, dotted lines are the calculations by CEM.
\begin{figure}[h]
\centering
\resizebox{6in}{5in}{\includegraphics{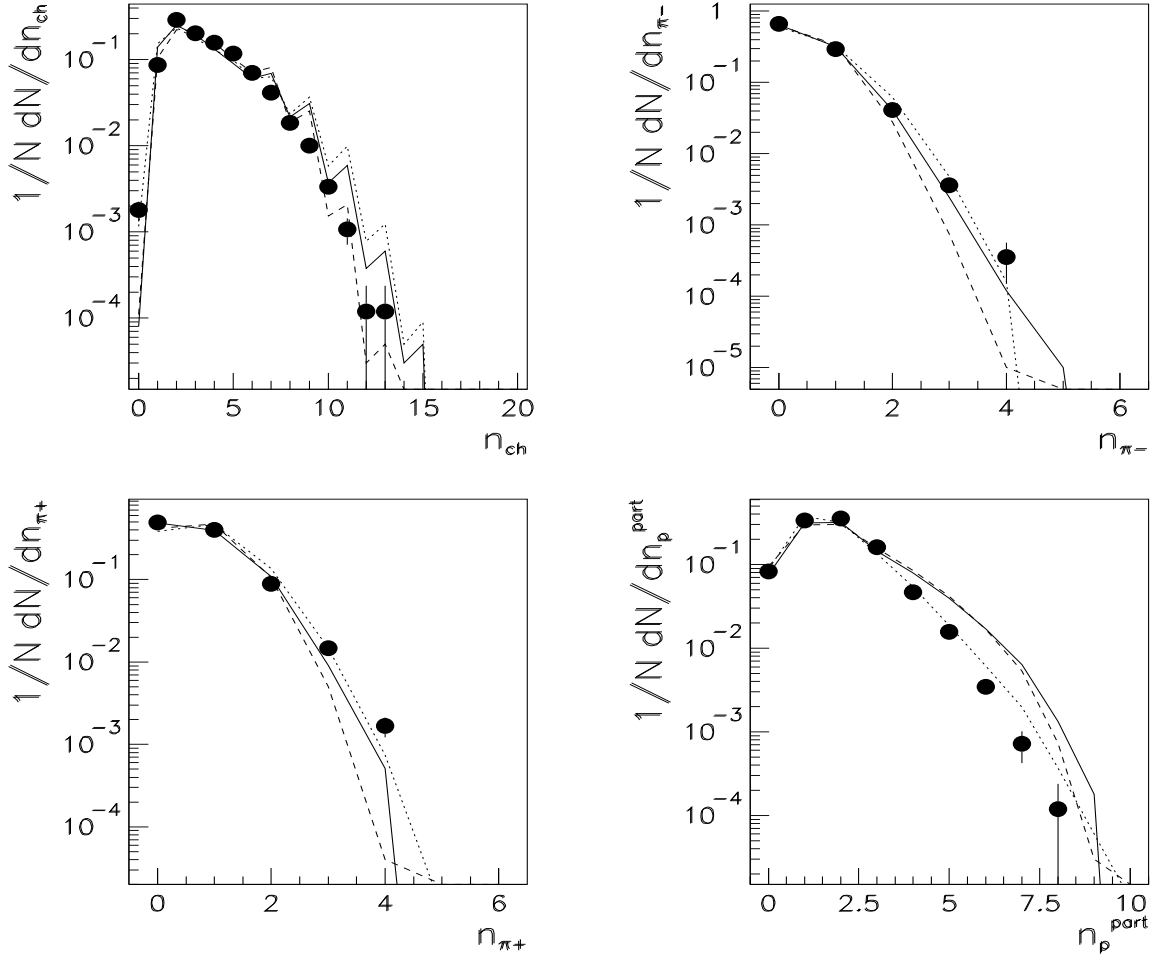}}
\caption{Multiplicity distributions in $p$C-interactions. Points are
the experimental data \protect \cite{pC}, lines are our calculations.}
\label{fig1}
\end{figure}
As seen, the models describe the distributions quite well.  The FRITIOF models
with and without $\Delta$-isobars (solid and dashed lines, respectively)
overestimates the proton participant multiplicity production.


Let us consider the average multiplicity dependence on the centrality
of the pC-interactions, presented in the table 1 and Fig. 2. One can
see, the average multiplicities of all charged particles,
$\pi^+$-mesons, proton-participants increase considerably passing from
peripheral interactions to the central ones. The average multiplicity
of $\pi^-$-meson changes slowly with increase of the parameter Q. The
highest value of $\pi^-$-meson multiplicity is in the events at Q=1
presented mainly $pn$-interactions. It is interesting, CEM that
describes well the multiplicity of $\pi^-$-mesons in all interactions
does not describe the dependence of this multiplicity on Q. At the same
time, CEM describes satisfactory the $\pi^+$-meson and
proton-participant multiplicities increase with enhance of Q. The
modified model FRITIOF also reproduces $\pi^-$, $\pi^+$ meson
multiplicities in all interactions. But the dependencies of $\pi^-$,
$\pi^+$ meson multiplicities on value of Q are not described by model
FRITIOF without $\Delta$-isobars. The modified FRITIOF model with
$\Delta$-baryons describes quantitative the $\pi^-$-meson multiplicity
at different Q (Fig. 2).
\begin{figure}[h]
\centering
\resizebox{6in}{5in}{\includegraphics{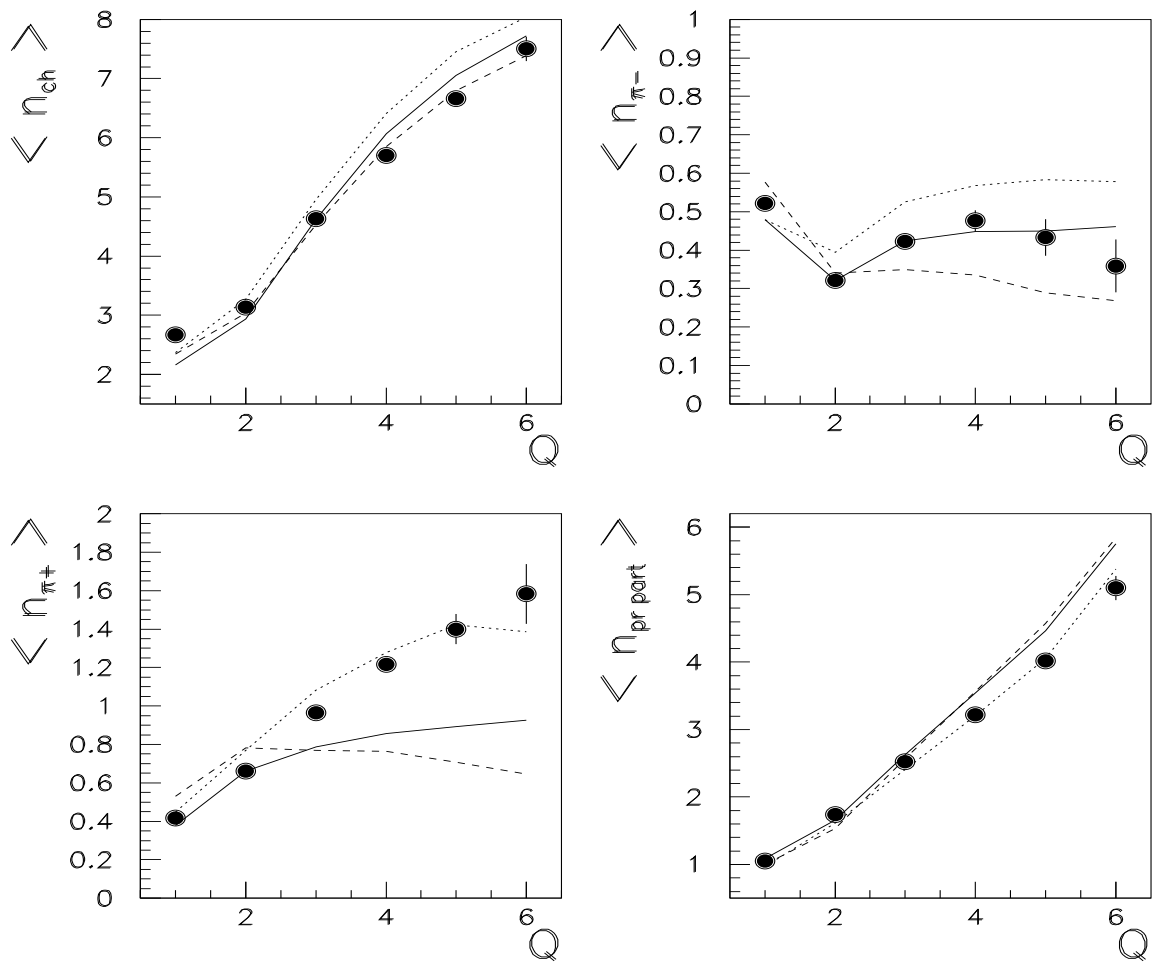}}
\caption{The average multiplicy dependences on $Q$.}
\label{fig2}
\end{figure}

In Fig. 3 the
average momenta, transverse momenta, and rapidities of $\pi^-$, $\pi^+$
mesons  are presented. The points are the experimental data of the table 2,
solid and dashed lines are the FRITIOF model calculations with and without
$\Delta$-isobars, dotted lines are CEM calculations, correspondingly.
As seen, the average momenta of pions decrease with increase of the collision
centrality. The theoretical models reproduce qualitatively the
dependence of the average momenta on the parameter Q. Perhaps, the
predictions of the model FRITIOF without $\Delta$-isobars are nearest
to the experimental values of the momenta and the transverse momenta of
pions. CEM and FRITIOF model with $\Delta$-isobar underestimate the average
momenta, the transverse momenta for all groups of the $p$C-events
subdivided by the value of the parameter Q. That is, they assume the
preferential production of pions with small momenta. The average values
of the polar angles of pions emission enhance with increase of Q. It
characterizes the process of pion production. The probability of pion
re-scatterings increases with decrease of the impact parameter, that
leads to the decrease of the average momenta and the increase of the
average emission angles of pion. This causes the weak
dependence of the average transverse momenta on the collision
centrality. The calculations by CEM and FRITIOF with $\Delta$-isobars
qualitatively reproduce the values of the average rapidities of pions
at different Q
\begin{figure}[h]
\centering
\resizebox{6in}{6in}{\includegraphics{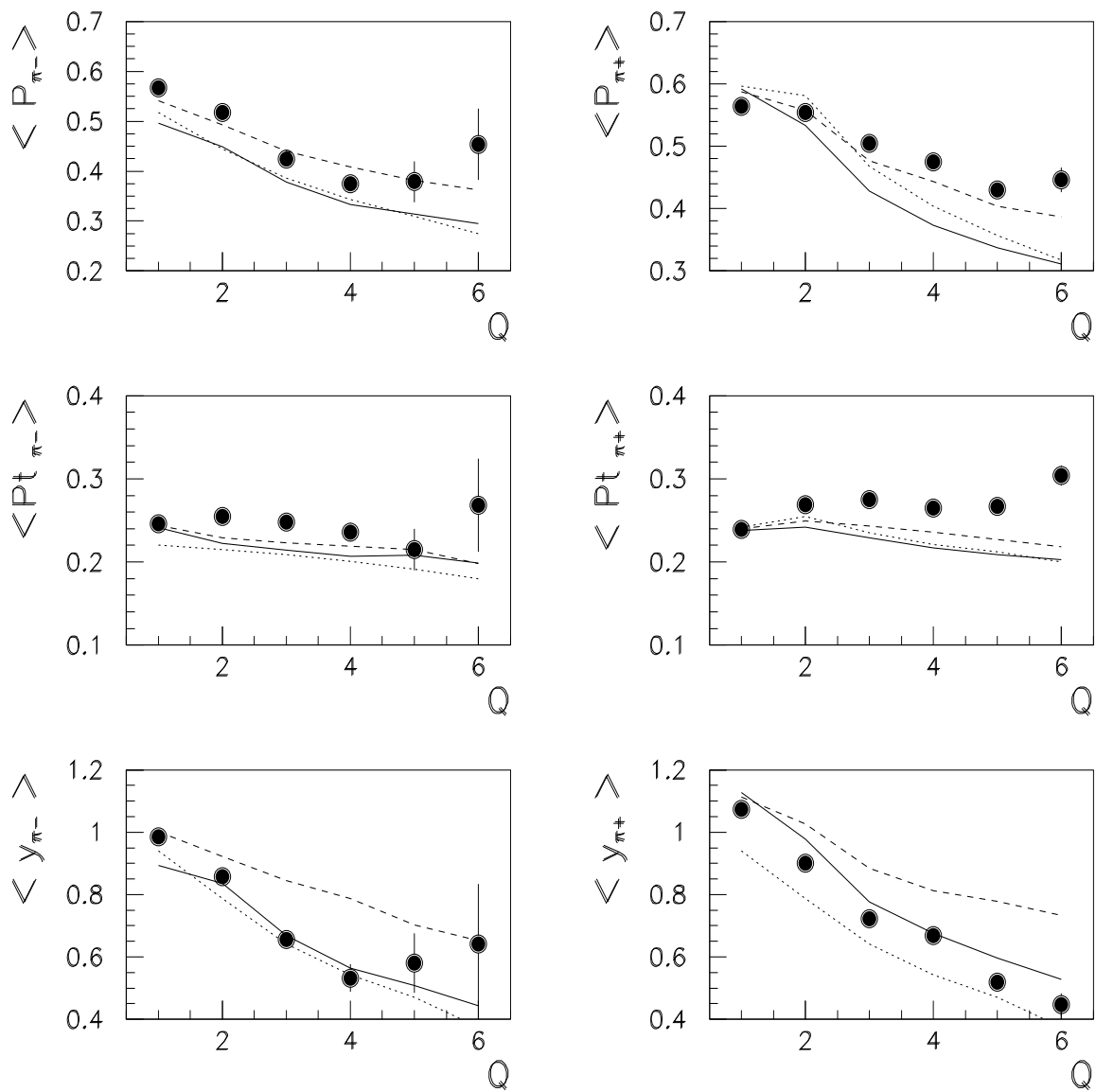}}
\label{fig3}
\caption{Dependencies of the average momenta, transverse momenta, and
rapidities of the produced particles on $Q$.}
\end{figure}

The differential distributions on momentum and on rapidity allow one
to obtain more specific conclusion about demerits of the applied
models. In Fig. 4, the momentum distributions of $\pi^-$-mesons
are presented in the six groups of the $p$C-events. The predictions and
the experimental data are distinguished at very small and large momenta.
Taking into account the large experimental errors in the region of large
momenta, one can consider the theoretical description as a satisfactory one.
Therefore the divergence of the average experimental and theoretical
values of momenta is connected with the small momentum region, mainly.
According to the Fig 4 (and also next Fig. 5) the CEM overestimates the
yield of the soft pions ($p<300$ MeV/c). The FRITIOF model without
$\Delta$-isobars underestimates considerable the production of the soft
pions. It explains the large values of the average momenta calculated by
this model. The small value of the average momenta of $\pi^-$-mesons in
the FRITIOF model with $\Delta$-isobars are caused by the insufficient
formation of hard pions. On the whole, the momentum distributions of
$\pi^-$-mesons in the separate groups of the $p$C-events are described
satisfactory by the models.
\begin{figure}[h]
\centering
\resizebox{6in}{2.9in}{\includegraphics{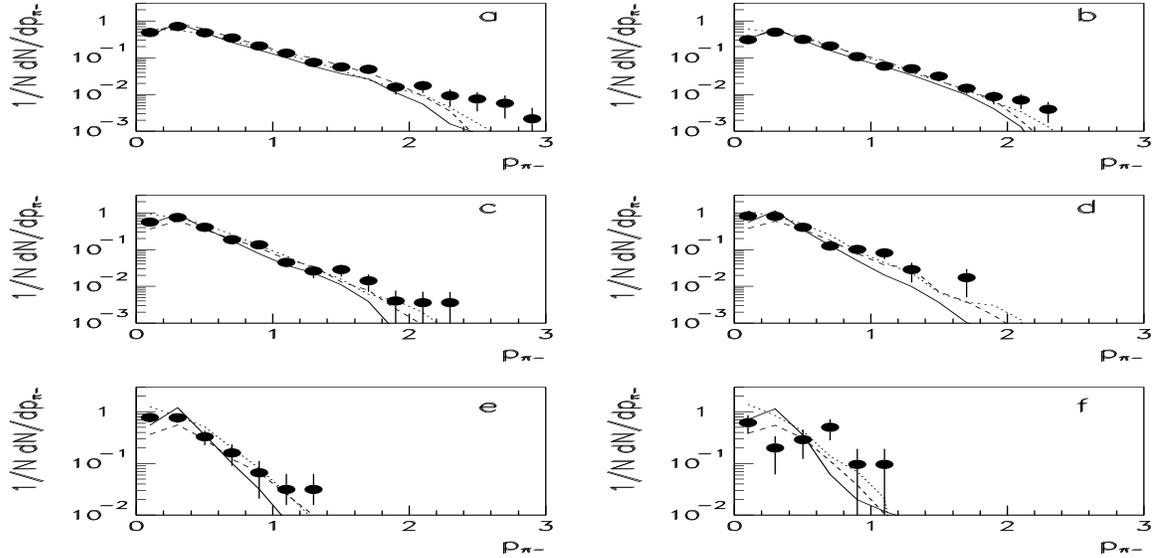}}
\caption{$\pi^-$-meson momentum distributions at $Q=$ 1, 2,
3, 4, 5 (figs. a -- e), and at $Q\geq 6$ (fig. f).}
\label{fig4}
\end{figure}

The situation gets complicated at analysis of the momentum
distributions of $\pi^+$-mesons in the groups of the $p$C-collisions
(Fig. 5). In the groups at Q=1, 2 presented mainly $pn$- and
$pp$-interactions, a good description of $\pi^+$-meson spectra is seen.
The models reproduce badly the experimental data in multi-nucleon collisions
at $Q\geq 3$. The study of such interactions can lead to a future
development of the FRITIOF model.
\begin{figure}[h]
\centering
\resizebox{6in}{2.9in}{\includegraphics{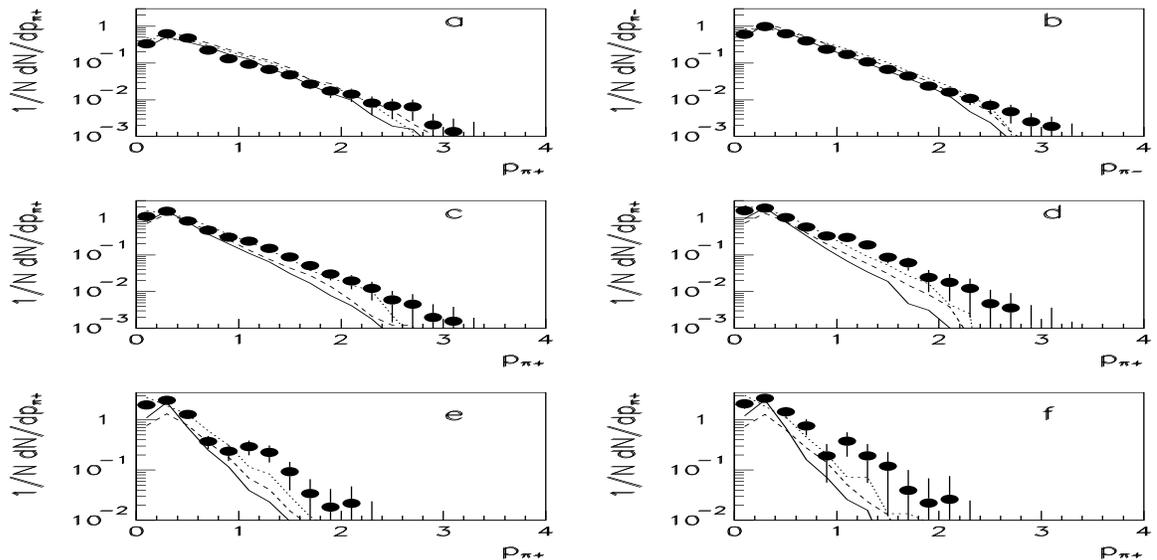}}
\caption{$\pi^+$-meson momentum distributions. Notation is the same as on the Fig. 4.}
\label{fig5}
\end{figure}

The rapidity distributions of $\pi^\pm$-mesons in the studied groups of the
$p$C-events give an interesting information about correlations between
theory and the experiment (Figs.~6,~7). As seen in the Fig. 6 (and also
next Fig. 7), the maximum of the y-distribution of pions moves to the
region of the carbon nucleus fragmentation with increase of Q. The
y-distribution of $\pi^-$-mesons in the events at Q=1 has two maxima at
$y \sim 0.5$ and $y\sim 1.5$. The two peaks structure are absent in the
next group. According to the Fig. 6a, the CEM underestimates
considerably the multiplicities of the fast $\pi^-$-mesons. The modified
FRITIOF model without $\Delta$-isobars overestimates the multiplicities
of $\pi^-$-mesons in the central region at $y\sim 1.1$. Moreover, it
underestimates the multiplicities of produced $\pi^-$-mesons in
multi-nucleon collisions (Fig. 6 c, d, e, f).
\begin{figure}[h]
\centering
\resizebox{6in}{5in}{\includegraphics{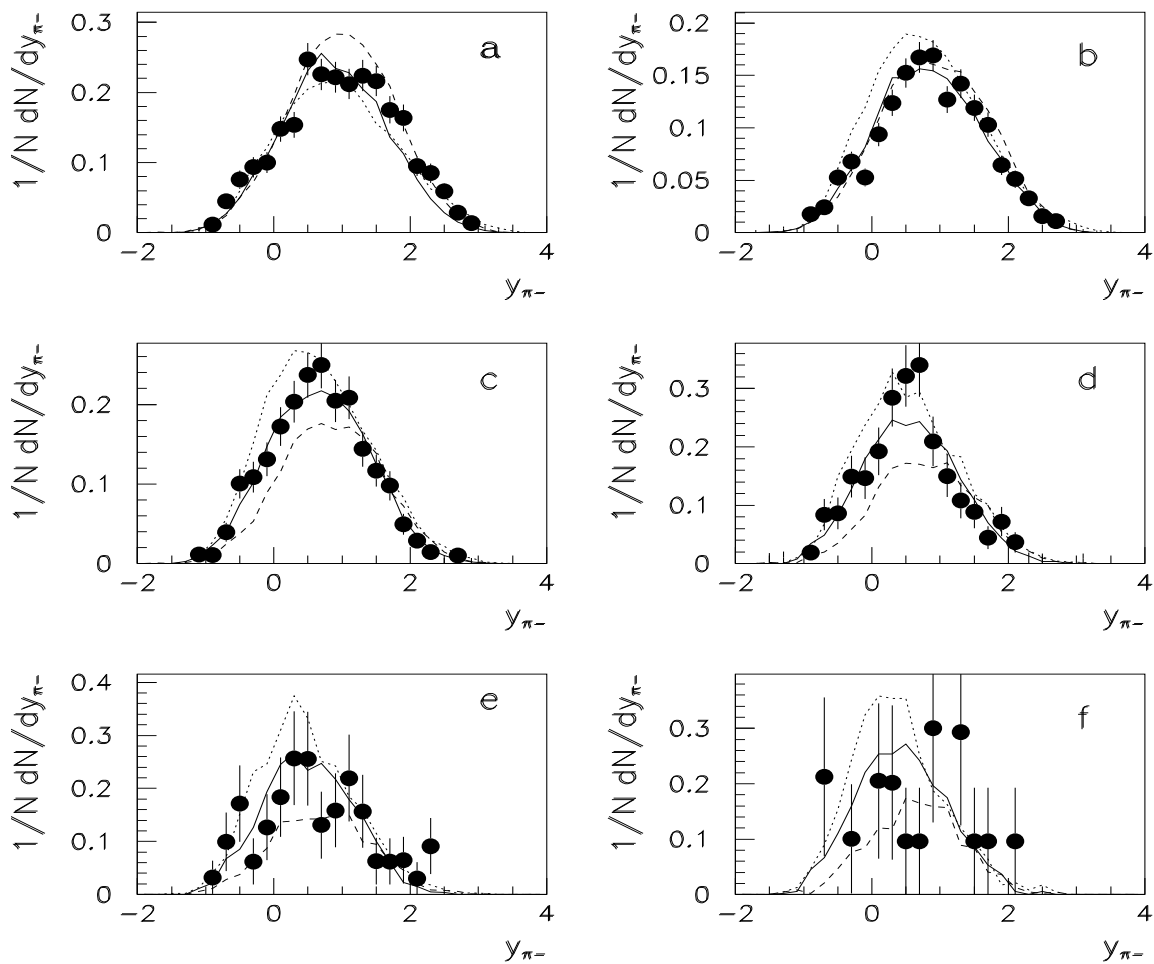}}
\caption{Rapidity distributions of $\pi^-$-mesons at different $Q$.
Notation is the same as on the previous figures.}
\label{fig6}
\end{figure}


In the group at Q=2 (Fig. 6b), CEM assumes the preferential production
of $\pi^-$-mesons in the fragmentation region of the target nucleus. It
takes place in the other groups of the $p$C-interactions (Fig. 6 c, d,
e, f). The calculations by the DFRITIOF model reproduce qualitatively
and quantitatively the experimental spectra of $\pi^-$-mesons (Fig. 6).

Let us turn to the rapidity distributions of $\pi^+$-mesons in the
$p$C-events at Q=1, 2, 3 (Fig. 7). At Q=1 CEM and the FRITIOF model with
$\Delta$-isobars underestimate the yield of $\pi^+$-mesons over a range
$0.5 < y <1.5$ (Fig. 7a). CEM gives a larger multiplicity of
$\pi^+$-mesons in the fragmentation region of the target nucleus as in
the case of $\pi^-$-mesons. The FRITIOF model without $\Delta$-isobars exceeds
the multiplicities at Q=1, 2 (Fig. 7a, 7b), and reduces the yield of
$\pi^+$-mesons in the multi-nucleon collisions (Fig. 7 c, d, e, f). Taking
into account $\Delta$-isobars in the FRITIOF model promotes
an insignificant increase of $\pi^+$-meson production in the
fragmentation region of the target nucleus. It is insufficiently for
the quantitative description of the experimental spectra of
$\pi^+$-mesons.
\begin{figure}[h]
\centering
\resizebox{6in}{5in}{\includegraphics{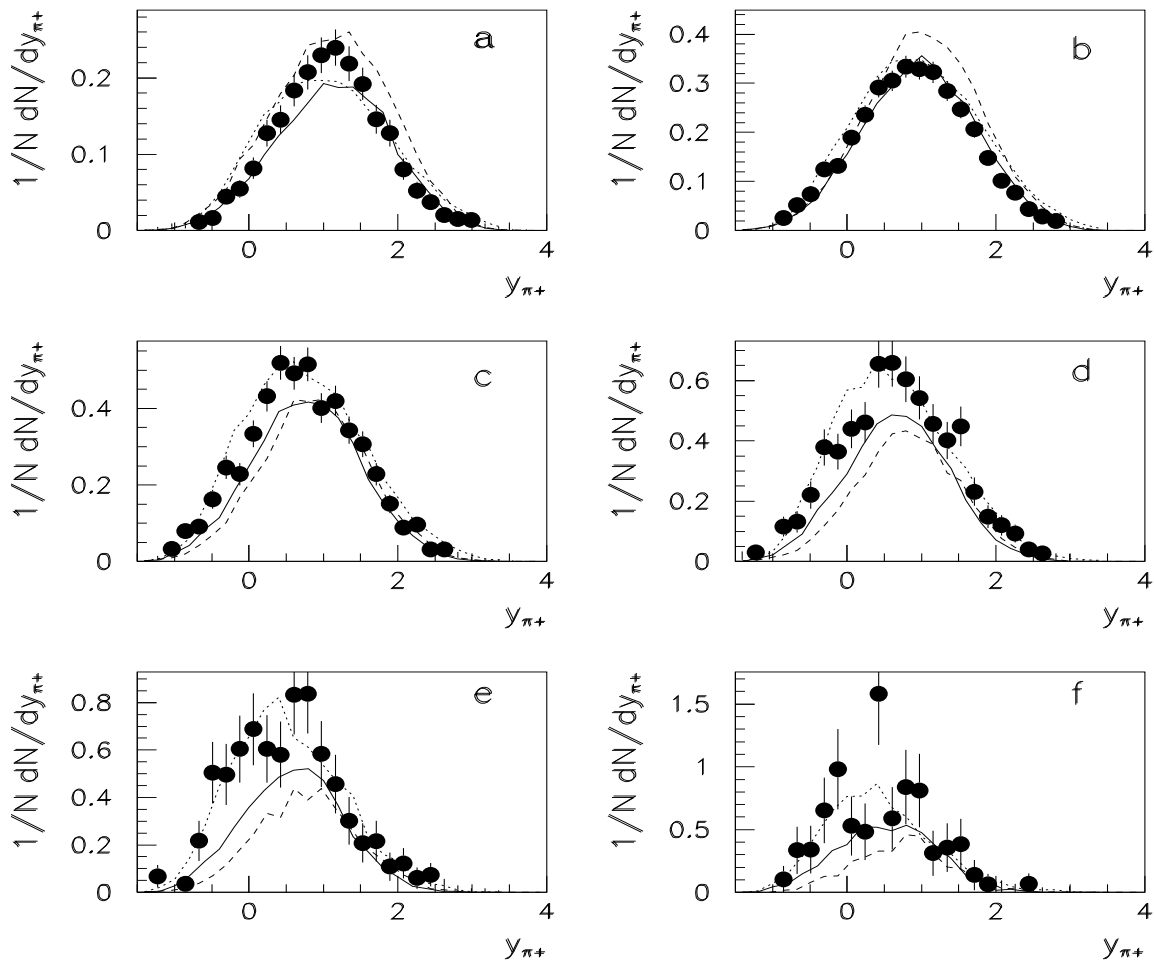}}
\caption{Rapidity distributions of $\pi^+$-mesons at different $Q$.
Notation is the same as on the previous figures.}
\label{fig7}
\end{figure}

The models reproduce better the average kinematical characteristics of
the participating protons in a dependence of the value Q. According to
the Fig. 8a, the spectra of the participating protons are
soften essential passing from the peripheral interactions to the
central ones. The average momentum of the protons decreases more than in
two times at changing of Q from 1 to 6. The change is connected mainly with
increase of the part of the target proton among total number of the
participating protons. The calculations show that at the average the
momentum of the target protons is less than 1.4 GeV/c.

To study the situation more carefully, the target protons were subdivided
into two groups: the first one
included protons at the momentum from 0.3 up to 0.75 GeV/c, the second
group contained ones at the momentum from 0.75 to 1.4 GeV/c. The
greater part of the target protons has been found in the first group. The weak
dependence of the average momentum on value Q is characteristic for
these protons (Fig. 8b). This fact is connected, perhaps, with the
small probability of inelastic scattering of the protons from the first
group. The average momentum of the fast target protons ($p>0.75$)
decreases with increase of Q. The average transverse momentum of the
participating protons has no dependent on Q starting from Q=2. It is
connected with a strong correlation between decrease of the average
momentum of the participating protons with increase of Q, and the growth
of their emission angles. This peculiarity is characteristic for the
target protons (as seen in the table 3) formed most of the
participating protons.
\begin{figure}[h]
\centering
\resizebox{6in}{6.2in}{\includegraphics{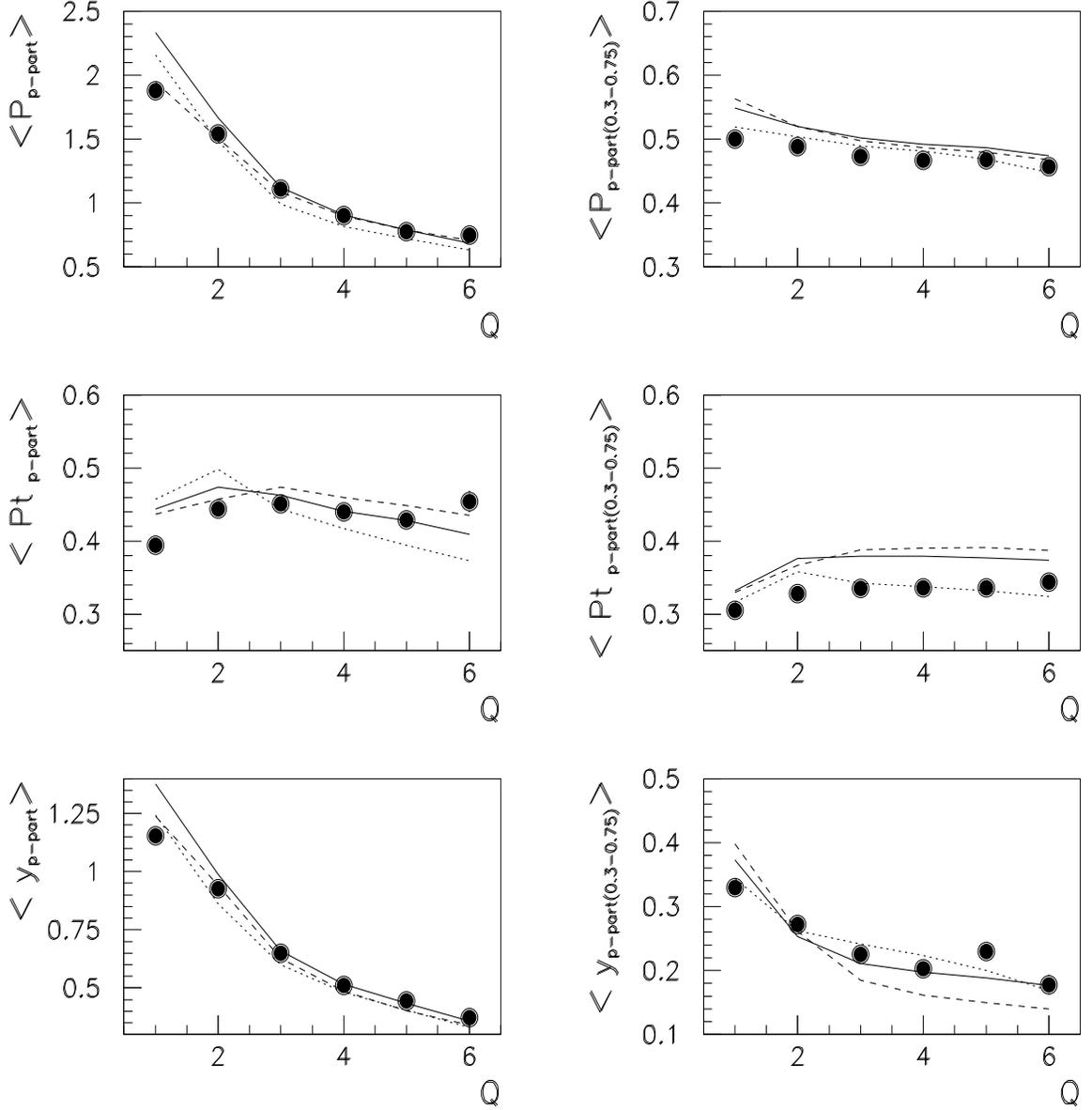}}
\caption{Average characteristics of the participating protons.}
\label{fig8}
\end{figure}

The leading protons ($p>1.4$ GeV/c) show (as seen in the table 4) quite
different dependence of the average transverse momentum on Q. Their
$p_T$ in the central interactions is in 1.2 -- 2 times higher than in
the peripheral ones. This effect does not influence practically on the
average $p_T$ of all participating protons due to small part of the
leading protons among them. The central interactions are marked
relatively small (25 \%) decrease of the momentum of the leading
protons in comparison with the peripheral interactions, but
considerably (2 -- 2.5 times) increase of the average emission angles
(table 4).
\begin{landscape}
\begin{table}
\caption {The average momenta and emission angles of the
leading and target protons in the $p$C-interactions at 4.2 GeV/c
at the different $Q$, e - experiment \protect \cite{pC}, m - the FRITIOF
model with $\Delta $- isobars.}
 \label{tabl4}
 \bigskip
 \hspace{-0.5cm}
{ 
\begin{center}
\begin{tabular}{|cc|c|c|c|c|c|c|c|}
\hline
 && & & & & & & \\
Q  &&  1  &  2  &  3  &  4  &  5  &  $\geq $6  & {\mbox All events} \\
 && & & & & & &
\\  \hline
 $<p_{p.part.}>$ (GeV/$c$) &e &  2.76$\pm$0.03 & 2.66$\pm$0.02 & 2.35$\pm$0.03 & 2.12$\pm$0.03
 & 2.02$\pm$0.06 & 2.02$\pm$0.10 & 2.58$\pm$0.01 \\
 $p\geq 1.4$ ~~~~(GeV/$c$) &m &
 2.894$\pm$0.006 & 2.643$\pm$0.05 & 2.277$\pm$0.006 & 2.113$\pm$0.007
 & 2.012$\pm$0.009 & 1.889$\pm$0.012 & 2.589$\pm$0.003
 \\ \hline
 $<p_t^{p.part.}>$ (GeV/$c$) &e &
 0.424$\pm$0.011 & 0.519$\pm$0.006 & 0.594$\pm$0.013 & 0.625$\pm$0.022
 & 0.682$\pm$0.049 & 0.816$\pm$0.076 & 0.519$\pm$0.005
 \\
 $p\geq 1.4$ ~~~~(GeV/$c$)    &m &
 0.453$\pm$0.002 & 0.498$\pm$0.002 & 0.533$\pm$0.003 & 0.509$\pm$0.004
 & 0.496$\pm$0.006 & 0.442$\pm$0.009 & 0.490$\pm$0.001
   \\ \hline
 $<\theta _{p.part.}>$ (grad) &e &
10.0$\pm$0.2 & 12.8$\pm$0.2 & 16.5$\pm$0.4 & 18.6$\pm$0.7 &
21.2$\pm$1.7 & 26.9$\pm$3.4 & 13.3$\pm$0.1
\\
$p\geq 1.4$ ~~~~(GeV/$c$)      &m &
 10.67$\pm$0.06 & 12.84$\pm$0.06 & 15.4$\pm$0.1 & 15.6$\pm$0.2 &
 15.7$\pm$0.2 & 14.5$\pm$0.3 & 12.94$\pm$0.04
 \\ \hline
 $<p_{p.part.}>$ ~~(GeV/$c$)     &e &
 0.764$\pm$0.008 & 0.717$\pm$0.004 & 0.665$\pm$0.005 &
 0.638$\pm$0.007 & 0.613$\pm$0.011 & 0.594$\pm$0.017 & 0.687$\pm$0.003
   \\
0.3$\leq p<$1.4 (GeV/$c$) &m &
 0.874$\pm$0.003 & 0.768$\pm$0.002 & 0.690$\pm$0.002 & 0.650$\pm$0.002
 & 0.621$\pm$0.002 & 0.586$\pm$0.002 & 0.692$\pm$0.001
   \\ \hline
$<p_t^{p.part.}>$ ~~(GeV/$c$)    &e &
0.357$\pm$0.005 & 0.388$\pm$0.002 & 0.400$\pm$0.004 & 0.400$\pm$0.005
& 0.396$\pm$0.008 & 0.410$\pm$0.012 & 0.391$\pm$0.002
\\
0.3$\leq p<$1.4 (GeV/$c$) &m &
 0.418$\pm$0.002 & 0.451$\pm$0.001 & 0.437$\pm$0.001 & 0.427$\pm$0.001
& 0.418$\pm$0.001 & 0.407$\pm$0.001 & 0.431$\pm$0.001
 \\ \hline
 $<\theta _{p.part.}>$ ~(grad)  &e &
 35.4$\pm$0.6 & 45.3$\pm$0.3 & 50.5$\pm$0.5 & 54.4$\pm$0.8 &
 52.7$\pm$1.2 & 59.8$\pm$2.1 & 47.1$\pm$0.2
 \\
0.3$\leq p<$1.4 (GeV/$c$) &m &
 33.2$\pm$0.2 & 46.4$\pm$0.2 & 52.9$\pm$0.2 & 55.7$\pm$0.2 &
 57.8$\pm$0.2 & 60.1$\pm$0.3 & 52.1$\pm$0.1
 \\ \hline
\end{tabular}
\end{center}
}
\end{table}
\end{landscape}

Going from the peripheral interactions enriched by NN-interactions
to the central ones, the average value of the rapidity of the
participating protons displaces from the value 1.1 to a smaller one. The
y-distributions of the participating protons (Fig. 9) show, that CEM
describes unsatisfactory $pn$-interactions (as seen in the group at Q=1
and y=1.5). The same minimum in the calculations is in the group at Q=2.
This minimum is caused by the poor reproduction of the proton spectra
of NN-collisions by CEM.
\begin{figure}[h]
\centering
\resizebox{6in}{5in}{\includegraphics{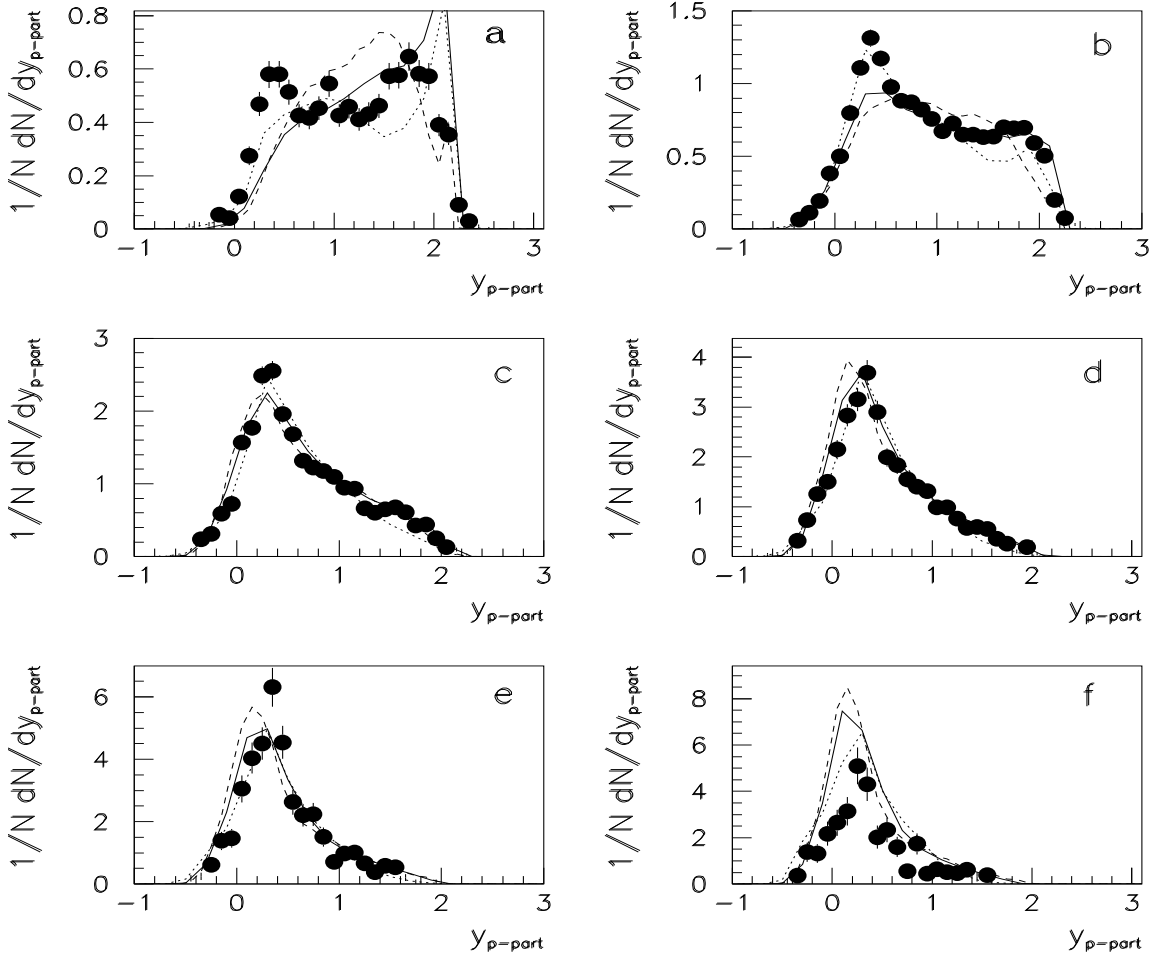}}
\caption{Rapidity distributions of the participating protons.
Notation is the same as on Fig. 4.}
\label{fig9}
\end{figure}

In the events at Q=1, 2 the experimental spectra of the participating
protons have a two peak structure. The wide peak at $y\sim 1.7$ is
defined by the leading protons (with $p>1.4$), and is similar to one existing
in $pn$-interactions. The peak at $y\sim 0.5$ is connected maybe
with the peak in the y-distributions of the $\pi^-$-mesons. It is
reflected the processes $n\rightarrow p+\pi^-$. The peak at $y\sim 0.5$
and Q=1 is not described by the models. However, CEM and DFRITIOF have
some better situation at a description of the peak at $y\sim 0.4$ and
Q=2. The elastic re-scattering of nucleons give the peak at Q=1 and
$y\sim 2$ in the calculations performed by CEM and FRITIOF.

The FRITIOF model without $\Delta$-isobars predicts an exceeding yield of
the protons in the central region at Q=1, 2 (see Fig. 9a, 9b). The
DFRITIOF model describes well the fast protons ($y>1$). The calculated
y-distributions are shifted to the fragmentation region of the target
nucleus with increasing of Q.
The DFRITIOF model and CEM reproduce qualitatively the y-distributions
of the protons in the $p$C-events at $Q>2$.

In the Fig. 10, the momentum
distribution of the participating protons in the six groups of the
$p$C-interactions are presented. The model calculations are in agreement
with the experimental data at $Q>2$ (Fig. 10 c, d, e, f).
\begin{figure}[h]
\centering
\resizebox{6in}{5in}{\includegraphics{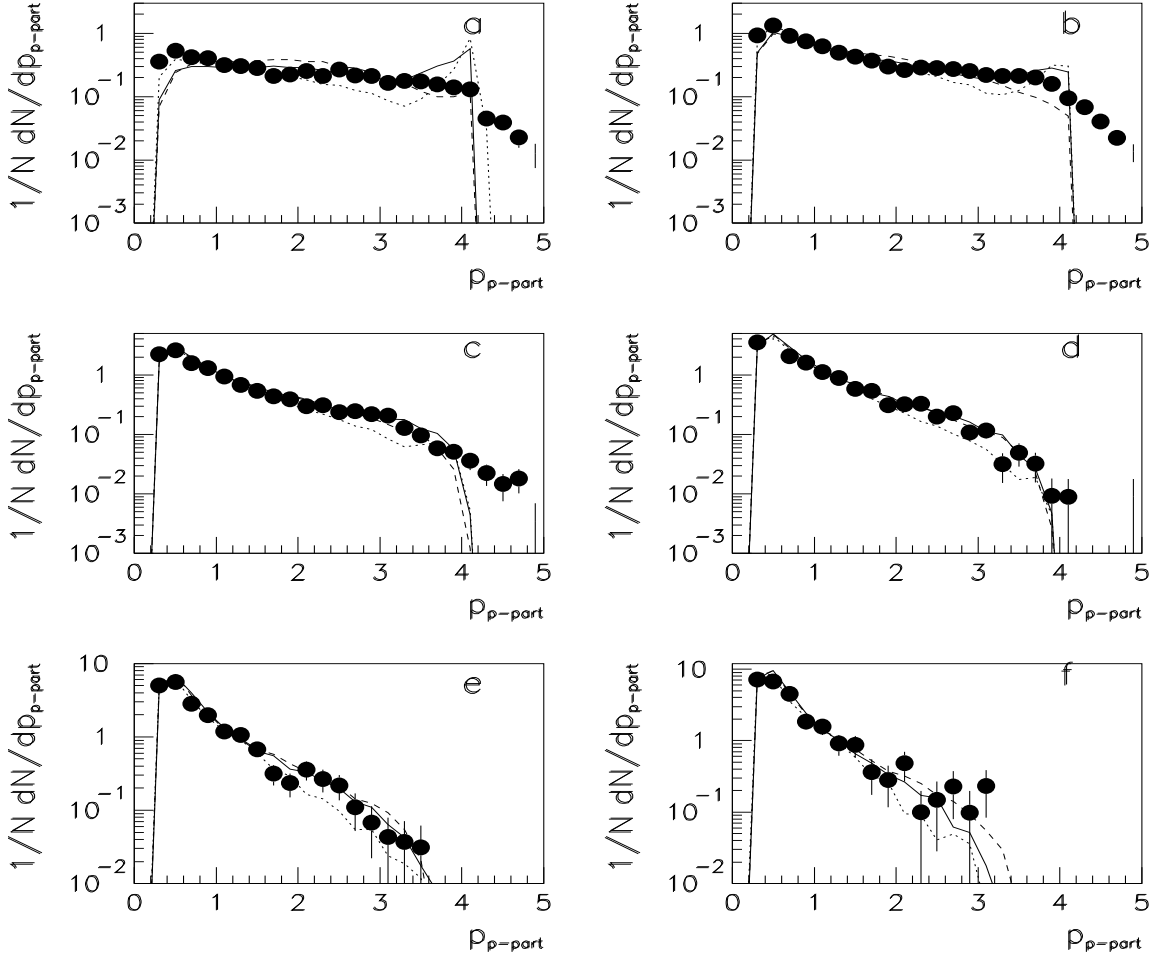}}
\caption{Momentum distributions of the participating protons.
Notation is the same as on Fig. 4.}
\label{fig10}
\end{figure}

The strong differences of the predictions are seen for the $p$C-events at
Q=1, 2 (related to the NN-interactions, basically). In the spectra calculated
by CEM the peak is observed at $p\sim 4$ GeV/c and Q=1
connected with the elastic re-scattering.
The model predicts a minimum at $p\sim 3$ GeV/c caused by unsatisfactory
description of the NN-interactions. In the calculations by FRITIOF model
without $\Delta$-isobars, this peak is absent. However, the model
predicts exceeding yield of the participating protons at the momentum
$\sim 2$ GeV/c and underestimates the production of the soft protons.
The predictions of the DFRITIOF model are near to the experimental
data, with the exception of the range $p\sim 4$ GeV/c.  Thus, we can conclude
 the used methods of the elastic re-scattering calculations are incorrect
(Fig. 10a).

Agreement of the experimental data and the calculations by the
DFRITIOF model is reached starting from $Q>2$. As before, the CEM
predicts the maximum at $p\sim 4$ GeV/c at Q=2.

\section*{Summary}
\begin{enumerate}

\item
The study of the $p$C-interactions in the dependence of the collision
centrality gives the important information about demerits of the
theoretical models. According to the presented data and calculations,
CEM describes unsatisfactory NN-collisions, related to the group of the
$p$C-events at Q=1, 2.

\item
For the first time, non-nucleonic degrees of freedom in nuclei
($\Delta^+$, $\Delta^-$-isobars) are taken into account in the FRITIOF
model. It allows one to improve the description of the dependence of
the $\pi^-$-mesons multiplicity on the collision centrality, and the
rapidity distributions of the secondary particles for all interactions
and for the groups of the $p$C-events at different Q.

\item
The improved FRITIOF model and CEM overestimate the elastic
re-scattering of the nucleons. We believe  for a correct description
of the experimental data it is needed to take into account the processes of the
diffraction dissociation of the nucleons in nuclei in the models.

V.V.Uzhinskii thanks RFBR(grand N 00-01-00307) and INTAS (grand N 00-00366)
for their financial support.

\end{enumerate}

\end{document}